\documentclass{article}
\usepackage{spconf,amsmath,graphicx}
\usepackage{amsfonts, amssymb}
\usepackage{tabularx, booktabs}
\usepackage{subcaption}  
\usepackage[justification=centering]{caption}
\usepackage{tabularx,booktabs}
\newcolumntype{Y}{>{\centering\arraybackslash}X}

\graphicspath{{}}

\usepackage{tikz,placeins}
\usepackage{multirow}
\usepackage{bbm}
\usepackage[hyphens]{url}
\usepackage{breakurl}
\usepackage[breaklinks]{hyperref}

\usetikzlibrary{arrows,decorations.markings}
\tikzstyle{every picture}+=[font=\rmfamily\it\bfseries\large]
\usetikzlibrary{positioning, shapes}
\pgfdeclarelayer{background}
\pgfdeclarelayer{foreground}
\pgfsetlayers{background,main,foreground}

\newcommand{\specialcell}[2][c]{%
  \begin{tabular}[#1]{@{}c@{}}#2\end{tabular}}

\title{Streaming end-to-end speech recognition with \\ jointly trained neural feature
enhancement}

\name{Chanwoo Kim, Abhinav Garg, Dhananjaya Gowda, Seongkyu Mun, and Changwoo Han}
\address{Samsung Research, Seoul, South Korea \\
  {\small \tt \{chanw.com, abhinav.garg, d.gowda, sk1213.mun, cw1105.han\}@samsung.com}}

\begin{document}
\ninept

\maketitle
\begin{abstract}
In this paper, we present a streaming end-to-end speech recognition model
based on Monotonic Chunkwise Attention (MoCha) jointly trained with enhancement
layers. Even though the MoCha attention enables streaming speech recognition 
with recognition accuracy comparable to a full attention-based approach, 
training this model is sensitive to various factors such as the difficulty of training
  examples, hyper-parameters, and so on. Because of these issues, 
speech recognition accuracy of a MoCha-based model for clean speech drops significantly
when a multi-style training approach is applied.
  Inspired by {\it Curriculum Learning} \cite{y_bengio_icml_2009_00}, 
  we introduce two training strategies: 
Gradual Application of Enhanced Features (GAEF) and Gradual Reduction of Enhanced Loss (GREL). With
GAEF, the model is initially trained using clean features. Subsequently, the portion of
outputs from the enhancement layers gradually increases. With GREL, the portion
  of the Mean Squared Error (MSE) loss for the enhanced output gradually reduces as training proceeds. In experimental results
on the LibriSpeech corpus and noisy far-field test sets, the 
  proposed model with GAEF-GREL training strategies
shows significantly better results than the conventional multi-style training
  approach.
\end{abstract}
\section{Introduction}
\label{sec:introduction}
 Since the introduction of end-to-end all neural speech recognition models
 \cite{j_chorowski_nips_2015_00}, there have been growing interests in these
 new models.
These models have significant advantages in structural simplicity
 and simplified inference to generate texts.
Compared to the conventional speech recognition system consisting of
multiple discrete components such as an Acoustic Model (AM), 
a Language Model (LM), a pronunciation dictionary,
and a decoder based on a Weighted Finite State Transducer (WFST),
a complete end-to-end all neural speech recognition system realizes  
all these functionalities using a single-structured neural network.
Further improvements in these end-to-end speech recognition systems
have been obtained thanks to a better choice of target units such as
Byte Pair Encoded (BPE) and {\it unigram language model} 
\cite{t_kudo_acl_2018_00} units, improved training methodologies, and so on. 
Another major advantage of these
end-to-end models is that they require much smaller memory footprint compared
to the conventional WFST-based models, which enables their widespread use for
on-device applications \cite{k_kim_asru_2019_00, a_garg_interspeech_2020_01}.

%
Attention-based Encoder-Decoder (AED) is perhaps the most well-known type of such
end-to-end speech recognition systems \cite{j_chorowski_nips_2015_00}. 
Recently, it has
been reported that the performance of the AED system outperforms the
conventional WFST-based decoders for large vocabulary speech recognition tasks
\cite{c_chiu_icassp_2018_00}. In spite of these achievements in performance, 
the biggest shortcoming of the AED-based speech recognition system is
its lack of streaming capability. To overcome this restriction, several
variations of AED approaches including  Monotonous Chunkwise Attention (MoCha) 
have been proposed \cite{c_chiu_iclr_2018_00}.
In our previous work \cite{k_kim_asru_2019_00}, we successfully employed MoCha
-based model for on-device dictation applications.

Recently, speech recognition has been widely adopted for AI speakers
\cite{c_kim_interspeech_2017_00} and home
appliance devices \cite{a_garg_interspeech_2020_01}. Therefore, far-field speech
recognition has become increasingly more important. It has been observed that
various kinds of data augmentation \cite{C_Kim_ASRU_2009_2} or approaches
motivated by auditory processing \cite{C_Kim_INTERSPEECH_2014_2} is especially helpful for far-field noisy environments.
On-the-fly data augmentation using an
 {\it acoustic simulator} \cite{c_kim_interspeech_2017_00,
c_kim_asru_2019_01} has been especially successful for these far-field speech
recognition scenarios. When
data augmentation using an  {\it acoustic simulator} is employed for
full-attention models, we observe that it even enhances performance for clean
utterances perhaps because of better regularization as will be shown in Sec.
\ref{sec:experimental_results}. However, the MoCha attention is generally 
unstable compared to the full attention \cite{h_miao_interspeech_2019_00}.
Therefore, it has been very difficult to obtain
good speech recognition accuracy if the same data augmentation strategy is directly
applied to MoCha training, as will be also shown in Sec.
\ref{sec:experimental_results}.

In this work, we place an enhancement block consisting of two layers of Long
Short-Term Memories (LSTMs)
in front of the streaming MoCha speech recognition model.  We refer this model
to as Neural Enhancement-Automatic Speech Recognition (NE-ASR). This combined 
model is jointly trained from scratch using random initialization without 
any needs to train each sub-model separately beforehand.
Even though there
have been many attempts 
to jointly optimize an enhancement block along with a speech recognition model, 
our approach is unique in the following two aspects:
\begin{itemize}
  \item Gradual Application of Enhanced Feature (GAEF)
  \item Gradual Reduction of Enhancement Loss (GREL)
\end{itemize}
To the best of our knowledge, our work is 
the first attempt in gradually reducing the enhancement loss in the joint
training of the entire models which has the net effect
of gradually combining two separate enhancement and encoder blocks
into a single combined encoder block.
As will be shown in Sec. \ref{sec:experimental_results}, this GREL strategy
significantly enhances speech recognition performance.
This NE-ASR model shows improved performance compared to
the baseline MoCha model for LibriSpeech \cite{v_panayotov_icassp_2015_00} {\tt test-clean} and 
{\tt test-other}, and
significantly better performance than the conventional data augmentation technique on
the same test set. For re-recorded far-field noisy test sets, our approach reduces Word
Error Rates (WERs) relatively by 55.93 \%  and  12.36 \% over the baseline
and the multi-style trained MoCha models, respectively.
\section{Related works}
\subsection{Monotonic Chunkwise Attention}
\label{sec:mocha}
%
%
\begin{figure*}[tbp]
  \begin{center}
    \resizebox{100mm}{!}{\input{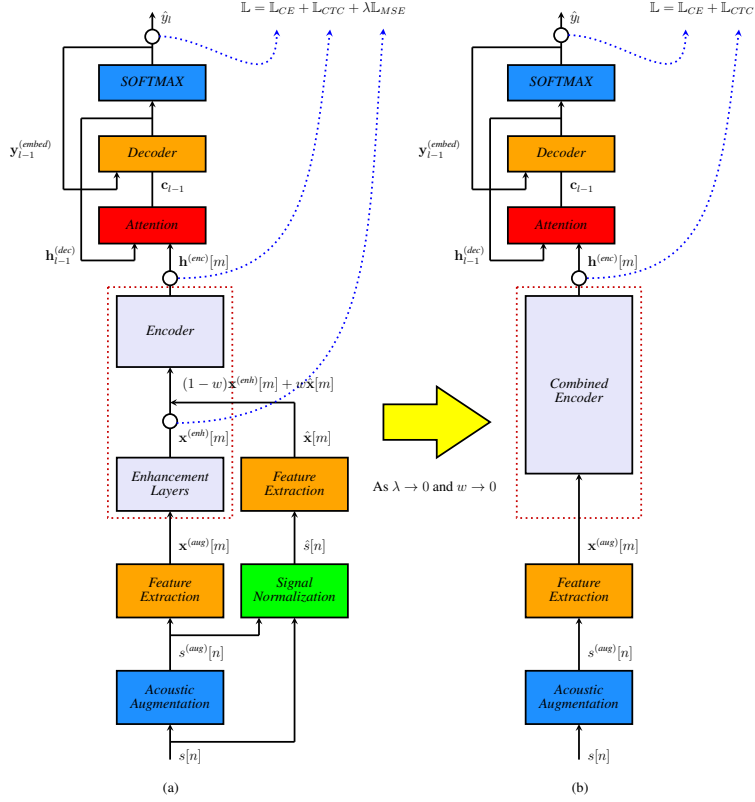}}
     \caption {
       \label{fig:training_strategy}
        The joint training strategy for a combination of neural enhancement and 
        an end-to-end speech recognition: The initial training pipeline is
        shown in Fig. \ref{fig:training_strategy} (a).
        $\lambda$ is a weight
        for the Mean Squared Error (MSE) loss $\mathbb{L}_{\text{MSE}}$ in
        \eqref{eq:joint_loss}.
        $w$ is a weight for the normalized clean feature $\hat{\mathbf{x}}[m]$ in 
        \eqref{eq:input_weighting}.
        As training proceeds, $\lambda$
        and $w$ reduce to zero, which transfigures the training pipeline into
        the one in Fig. \ref{fig:training_strategy} (b).
     }
  \end{center}
\vspace{-7mm}
\end{figure*}
In this section, we briefly describe the Monotonic Chunkwise Attention (MoCha)
algorithm \cite{c_chiu_iclr_2018_00}. 
In the MoCha model, there
are two attention mechanisms: a hard monotonic attention followed by a soft chunkwise
attention. The hard monotonic attention is employed to determine which element 
should be attended from a sequence of hidden encoder outputs
$\mathbf{h}^{\text{({\it enc})}}[m]$.
The hard monotonic attention is obtained from the hidden encoder output 
$\mathbf{h}^{\text{({\it enc})}}[m]$ at the frame index $m$ and the hidden
decoder output $\mathbf{h}^{(\text{\it dec})}_{l-1}$ at the output label index
$l-1$ as follows:
\begin{subequations}
  \begin{align}
    e^{\text{({\it mono})}}_l[m] & = \text{\it MonotonicEnergy} 
      (\mathbf{h}^{(\text{\it enc})}[m], \mathbf{h}^{(\text{\it dec})}_{l-1}) \\
    a^{\text{({\it mono})}}_l[m] & = \sigma(e^{\text{(\it mono)}}_l[m]) \\
    z_l[m] & \sim \text{\it Bernoullli}(a^{(\text{\it mono})}_l[m])
    \label{eq:hard_output}
  \end{align}
\end{subequations}
where $\sigma(\cdot)$ is a logistic sigmoid function and {\it MonotonicEnergy} is the 
energy function defined as follows \cite{c_chiu_iclr_2018_00}:
\begin{align}
  \text{\it MonotonicEnergy}&(\mathbf{h}^ {\text{({\it enc})}}[m],\;\mathbf{h}^{\text{({\it dec})}}_{l-1})
  \nonumber \\
  \;  = g \frac{v^{\intercal}}{\| v \|} & \tanh (W^{\text{({\it dec})}}
  \mathbf{h}^{\text{({\it dec})}}_{l-1} + W^{\text{({\it enc})}}  \mathbf{h}^ {\text{({\it enc})}}[m] + b ) + r
\end{align}
where $v, W^{\text{({\it dec})}}, W^{\text{({\it enc})}}, b, g$, and $r$ are learnable variables. After finding out the
position to attend using \eqref{eq:hard_output}, a soft attention with a fixed
chunk size is employed to generate the context vector that will be given as an
input to the decoder.
For all model training in this paper, an Adam optimizer
\cite{d_p_kingma_iclr_2015_00} is employed. We observe 
that careful learning rate scheduling is required to train high-performing MoCha
models. In this work, we employ a pre-training stage similar to that in
our previous work \cite{k_kim_asru_2019_00}. During the pre-training stage,
when a new layer is added or the pool size of a max-pooling layer is changed,
the learning rate is lowered to 1.0e-4 and linearly increased to 3.0e-4.
In LibriSpeech training described in Sec. \ref{sec:experimental_results}, 
this pre-training stage continues up to 2.25 epochs.
After the pre-training stage, the learning rate is maintained at 3.0e-4 up to
the eight full epochs. After finishing eight epochs of training, the learning
rate is reduced using the Newbob method. Without careful learning rate control,
MoCha model does not converge well especially when far-field data augmentation is
applied to the training set.
\subsection{Data augmentation using an acoustic simulator}
\label{sec:data_augmentation}
The {\it acoustic simulator} we used in our joint-training strategy is based on
the structure described in our earlier works 
\cite{c_kim_interspeech_2017_00, c_kim_asru_2019_01,
c_kim_interspeech_2018_00}.
Our {\it acoustic simulator} is designed to simulate the general case
with $I$ sound sources and $J$ microphones in a cuboid-shaped room with 
acoustically reflective walls \cite{c_kim_interspeech_2018_00}.
However, for brevity, we consider a single microphone case in this paper. The
relationship between the sound source $x_0[n]$ and the microphone output
$s^{\text{({\it aug})}}[n]$ is given by the following equation:
\begin{align}
  s^{\text{({\it aug})}}[n] = h_0[n] * x_0[n]  + \sum_{i=1}^{I-1} \alpha_i h_i[n]
  * x_i[n],
    \label{eq:rir_eq}
\end{align}
where  $x_i[n]$ is the signal from the $i$-th source. $\alpha_i$ and $h_i[n]$ 
are the scaling coefficient and the room impulse response associated with each
sound source.
When the {\it image method} is employed, the room impulse response $h_i[n]$ is given by 
the following equation
\cite{c_kim_interspeech_2018_00}:
\begin{align}
  h_i[n] = \sum_{k = 0}^{\infty} \frac{r^{g_{i, k}}}{d_{i, k}}
    \delta \left[n -\left \lceil{\frac{d_{i, k} f_s}{c_0}}\right \rceil \right],
      \label{eq:h_n_calculation}
\end{align}
where $d_{i,k}, r, g_{i,k}, f_s$, and $c_0$ are the distance between the
microphone and the sound source image, the reflection coefficient of the wall, the number of
reflections in the acoustic path connecting the microphone and the sound source
image, the sampling rate of a signal, and the speed of sound in the air, respectively.
In this work, we employ a similar configuration to our previous work
\cite{c_kim_interspeech_2017_00} regarding the distribution
of room sizes, reverberation times ($T_{60}$) , Signal-to-Noise Ratios in
Decibel (SNR dB), sound source locations and so on.
\section{Joint-training strategy}
\subsection{Neural Enhancement - Automatic Speech Recognition (NE-ASR) model}
In this section, we explain our joint training strategy of enhancement and
speech recognition models in detail.  The entire structure of the Neural Enhancement and 
Automatic Speech Recognition (NE-ASR) model is shown in Fig. 
\ref{fig:training_strategy}. This NE-ASR model structure does not change during the
training phase. However, since we gradually reduce the weight for the
enhancement loss $\lambda$ and the weight $w$ for the normalized  clean feature
to zero as training proceeds, the structure is finally equivalent to the
model in Fig. \ref{fig:training_strategy} (b).

$s[n]$ at the bottom of Fig. \ref{fig:training_strategy} is the
clean speech signal with $n$ being the sample index. Using the {\it acoustic simulator} 
described in Sec. \ref{sec:data_augmentation} \cite{c_kim_interspeech_2017_00,
c_kim_interspeech_2019_00}, we obtain simulated far-field noisy utterances
$s^{\text{({\it aug})}}[n]$. For {\it on-the-fly} data augmentation, we use
the {\it example server} system described in \cite{c_kim_asru_2019_01}.
As will be discussed in Sec. \ref{sec:delay_energy_norm}, 
there is a mismatch in time-delay and energy between the augmented signal
$s^{\text{({\it aug})}}[n]$
and the original signal $s[n]$. This mismatch is compensated using the
Delay-Energy Normalization (DEN) block. 
This DEN block is introduced because we believe that the neural network cannot
be trained effectively to learn the mapping from noisy features to clean
features if the time delay or the energy level difference vary randomly.

The output from the DEN block is
represented by $\hat{s}[n]$ in Fig. \ref{fig:training_strategy}.
 We use {\it power-mel} features of the order of 40
\cite{c_kim_interspeech_2019_00, c_kim_asru_2019_00}, which is motivated by the
power-law nonlinearity with a power coefficient of $1/15$
\cite{c_kim_taslp_2016_00}. We use a window length of $25$ {\it ms} with the
period between successive windows of $10$ {\it ms}. 
We use  {\it power-mel} features instead of the
more widely used {\it log-mel} features since {\it power-mel} features have
shown better performance in our previous studies
\cite{c_kim_interspeech_2019_00, c_kim_asru_2019_00}. We obtain
{\it power-mel} features $\mathbf{x}^{\text{({\it aug})}}[m]$ and $\hat{\mathbf{x}}[m]$ from
the augmented and the normalized clean speech signals respectively with $m$
being the frame index. The enhancement layers in Fig. \ref{fig:training_strategy} 
consist of two layers of LSTM. The unit sizes of these two layers are 1024 and
40 respectively. The enhanced feature $\mathbf{x}^{{\text{({\it enh})}}}[m]$ 
is obtained as the output of the enhanced layers, and the linear 
combination of the enhanced feature $\mathbf{x}^{{\text{({\it enh})}}}[m]$
and the normalized clean feature $\hat{\mathbf{x}}[m]$
is used as the input to the MoCha based streaming end-to-end speech recognition
model. This feature combination will be described in detail in the following
Sec. \ref{sec:gaef}. 
In the MoCha model, the encoder consists of six layers of
uni-directional LSTMs with the unit size of 1536
that are interleaved with 2:1 max-pooling layers 
in the lower three layers as  in \cite{k_kim_asru_2019_00}. 
Thus, the overall temporal reduction factor is 8:1.
\subsection{Gradual application of enhanced features (GAEF)}
\label{sec:gaef}

In machine learning, it has been frequently observed that models are trained
more effectively if they are trained with easier examples in the early
stage of training \cite{y_bengio_icml_2009_00}. Motivated by this observation, we 
start training the NE-ASR model with easier examples $\hat{\mathbf{x}}[m]$.
In the initial stage of training, the output from the enhancement block 
$\mathbf{x}^{\text{({\it enh})}}[m]$ might be still noisy, since this
enhancement layers are not sufficiently trained.
The input to the MoCha speech recognition model 
$\mathbf{x}^{\text{({\it comb})}}[m] $ is a linear combination of the
enhanced feature $\mathbf{x}^{\text{({\it enh})}}[m]$ and the normalized clean feature
$\hat{\mathbf{x}}[m]$ that is given by the following equation:
\begin{align}
  \mathbf{x}^{\text{({\it comb})}}[m] & = (1-w) \mathbf{x}^{\text{({\it
  enh})}}[m] + w \hat{\mathbf{x}}[m],  \label{eq:input_weighting}  \nonumber \\
      w  :  \;\; & 1.0 \rightarrow  0.0 \;\; \text{as training proceeds}.
\end{align}
where $w$ is the weighting coefficient. As the training proceeds, the $w$
decreases linearly from one to zero. In our experiments with {\it LibriSpeech}
training in Sec. \ref{sec:experimental_results}, $w$ becomes zero after eight full epochs of training.
\subsection{Gradual reduction of enhancement loss (GREL)}
\label{sec:grel}
The loss function $\mathbb{L}$ that we used in this joint training is a combination of the
Cross-Entropy (CE) loss $\mathbb{L}_{\text{\it CE}}$, 
Connectionist Temporal Classification (CTC) loss $\mathbb{L}_{\text{\it CTC}}$
\cite{a_graves_icml_2006_00} and the Mean Squared Error (MSE) loss, which is 
given by the following equation:
\begin{align}
  \mathbb{L} &  = \mathbb{L}_{\text{\it CE}}  + \mathbb{L}_{\text{\it CTC}} 
              + \lambda \mathbb{L}_{\text{\it MSE}},  \label{eq:joint_loss} \\
      \lambda  :  \;\; & 1.0 \rightarrow  0.0 \;\; \text{as training proceeds}.
      \nonumber
\end{align}
The CTC loss $\mathbb{L}_{\text{\it CTC}}$ and the CE loss $\mathbb{L}_{\text{\it CE}}$ 
are computed from the softmax outputs from the encoder and the decoder
respectively, as shown in Fig.
\ref{fig:training_strategy} \cite{k_kim_asru_2019_00, s_kim_icassp_2017_00}.
The MSE loss is computed from the enhancement layer output
$\mathbf{x}^{\text{({\it enh})}}[m]$
using the normalized clean feature
$\hat{\mathbf{x}}[m]$ as the target. Similar to the case of GAEF in Sec.
\ref{sec:gaef}, in our experiments with LibriSpeech training in
Sec. \ref{sec:experimental_results}, $\lambda$ value is one at the
beginning of training and linearly reduces to zero after finishing eight epochs of
training.
\subsection{Delay-Energy Normalization (DEN)}
  \label{sec:delay_energy_norm}
In this section, we describe {\it delay-energy normalization} employed to
obtain the normalized speech signal $\hat{s}[n]$ in 
Fig. \ref{fig:training_strategy}.  
From \eqref{eq:rir_eq}, assuming that $x_0[n] = s[n]$ is the target sound source and 
the remaining sound sources $x_i[n], \; 1 \le i \le I-1$ are noise sound sources, 
this equation expressed as follows:
\begin{align}
  s^{\text{({\it aug})}}[n] = h_0[n] * s[n] + \nu[n],  \label{eq:simplied_y_n}
\end{align}
where $\nu[n]$ is the combined noise-terms in \eqref{eq:rir_eq} defined by 
$\sum_{i=1}^{I-1} \alpha_i h_i[n] * x_i[n]$. From \eqref{eq:simplied_y_n},
we observe that $s^{\text{({\it aug})}}[n]$ has delays compared to the target signal $x_0[n]$ 
because of the room impulse response $h_0[n]$. 
From \eqref{eq:h_n_calculation}, we observe
that the first arriving impulse response is delayed by $\frac{f_s \times d}{c_0}$.
Thus, in the DEN block, this amount of delay component is added to the clean
speech.
The energy of $y[n]$ also may be very different from that of $s[n]$.
We measure the 95-percentile of frame energies of $s^{\text{({\it aug})}}[n]$, 
and $s[n]$ is scaled to have the same 95-percentile of frame energies.

\begin{table*}[!htbp]
  \renewcommand{\arraystretch}{1.3}
  \centering
        \caption{\label{tbl:ne_asr_result}
        Word Error Rates (WERs) using three different neural network models on
        the LibriSpeech corpus and the far-field noisy test set. If a specific
        training strategy is employed, it is marked by the {\tt O} symbol.
        Otherwise, it is marked by the {\tt X} symbol.
  \vspace{-2mm}
        }
  \begin{minipage}{0.9\linewidth}

\smallskip
    \begin{tabularx}{\textwidth}{  c * 7 {Y}}
\toprule
    \multirow{2}{*}{Model Type} & 
    \multicolumn{3}{c}{Training Strategy} &
    \multicolumn{3}{c}{Noise Types} \\
    \cmidrule{2-4} \cmidrule{5-7} 
    &
    \specialcell{Far-field\\Data Augmentation} &
    GAEF &
    GREL &
      \specialcell{LibriSpeech \\{\tt test-clean}} & 
      \specialcell{LibriSpeech \\{\tt test-other}} & 
      \specialcell{Far-Field \\ Noisy }\\
\midrule
      \multirow{2}{*}{\specialcell{BLSTM with Full Attention \\(1024 LSTM unit
      size)}}  & {\tt X } &  -  &  - &  4.19   \% &  13.95 \% &  70.22 \% \\
                            & {\tt O } & -  & - &  4.16 \% &  11.44 \% &  30.33 \%  \\
\hline
      \multirow{2}{*}{\specialcell{ULSTM with MoCha \\ (1536 LSTM unit size)}}
      & {\tt X } & -  &  - &  6.06 \% &  17.90 \% &  78.83 \% \\
                            & {\tt O } & -  & - & 7.10 \% &  16.42 \% &  39.64 \%\\
\hline
      \multirow{4}{*}{\specialcell{NE-ASR \\ without DEN}}
        &  {\tt O} & {\tt X}& {\tt X} & 8.01 \% & 19.43 \% & 43.79 \% \\
        &  {\tt O} & {\tt O}& {\tt X} & 6.99 \% & 19.35 \% & 45.61  \% \\
    
        & {\tt O} & {\tt X} & {\tt O}  & 6.86 \% & 16.28 \% & 36.73 \% \\

        & {\tt O} & {\tt O} & {\tt O} & 5.98 \%
        & 15.70 \% &   35.35  \% \\
\hline
      {\specialcell{NE-ASR \\ with DEN}}
   
        & {\tt O} & {\tt O} & {\tt O} & \textcolor{blue}{\bf 5.79} \%
        & \textcolor{blue}{\bf 15.63} \% &  \textcolor{blue} {\bf 34.74 } \% \\
\bottomrule
\end{tabularx}
  \end{minipage}
  \vspace{-5mm}
\end{table*}
\section{Experimental Results}
\label{sec:experimental_results}
In this section, we present speech recognition results 
obtained using the NE-ASR model on the LibriSpeech database 
\cite{v_panayotov_icassp_2015_00} and in-house far-field noisy test set.
For training, we used the entire 960 hours LibriSpeech training
set consisting of 281,241 utterances. For evaluation, we used the
official 5.4 hours {\tt test-clean} and 5.1 hours {\tt test-other} sets.
To evaluate the performance on far-field noisy environments, re-recording was
done by playing back 100 command utterances using a loud speaker at 5-meter distance
in a real room. 
The reverberation time in this room was measured to be $T_{60}=430$ {\it ms}. 
We simulated far-field additive noise by playing back four different types of
noise using loud speakers: babble, music, two different types of sound from a television.
We created noisy utterances at five different SNR dB levels: 0-dB, 5-dB, 10-dB, 15-dB, and 20-dB.
This far-field noisy test set was previously used in
our work described in \cite{c_kim_asru_2019_01}.

Table \ref{tbl:ne_asr_result} shows the speech recognition experimental results using
three different models. 
To implement and train this model, we use the
{\tt Keras} framework  \cite{f_chollet_keras_2015_00} with the
{\tt Tensorflow} 2.3 toolkit. 
For all the cases in Table \ref{tbl:ne_asr_result}, the models are trained
twice up to 25 
full epochs and these two WERs are averaged. During the inference for
evaluation, we use the beam size of 12. 
The BFA stands for Bidirectional LSTMs (BLSTMs) in the encoder with a Full Attention (BFA), 
which has a similar structure to those our previous work 
\cite{c_kim_interspeech_2019_00, c_kim_asru_2019_00}. For
MoCha and NE-ASR models, we use Unidirectional LSTMs (ULSTMs) with the
MoCha attention for streaming speech recognition. The unit sizes of
the BLSTM and ULSTM in the encoder is 1024 and 1536, respectively. For all these three models, a single layer
of ULSTM of a unit size of 1000 is employed for the decoder part. 
As shown in this Table, in the case of BFA, the performance for all these three
test sets improve by applying {\it on-the-fly} data augmentation using the
{\it acoustic simulator} \cite{c_kim_asru_2019_01}. We note that augmentation
using noisy utterances even enhances performance on the {\tt test-clean} test set
perhaps because of better regularization. However in the case of the MoCha model,
after applying the same data augmentation, the performance on the {\tt test-clean} test set
degrades by 14.65 \% relatively even with a very careful learning rate schedule
mentioned in Sec. \ref{sec:mocha}. Without careful learning rate scheduling,
the model even does not converge. In the case of the NE-ASR model, GREL strategy
significantly improves the performance compared to the MoCha model with
the far-field data augmentation. GAEF itself does not show performance
improvement when it is employed without GREL. However, when we employ both of
the GAEF and GREL algorithms, the best performance is achieved. As shown in
Table \ref{tbl:ne_asr_result}, the DEN approach brings further improvement.
Overall, the NE-ASR model with the GAEF-GREL training strategy
shows significantly better results than the conventional data augmentation
technique with relative improvements of 18.45 \%, 4.81 \%, and 12.36 \%
on the LibriSpeech {\tt test-clean}, {\tt test-other} and the re-recorded
far-field noisy test sets respectively.
\section{Conclusions}
In this paper, we present a Neural Enhancement-Automatic Speech Recognition
(NE-ASR) model that consists of an enhancement model and an end-to-end speech
recognition model. These two models are jointly trained from scratch using random
initialization without pre-training each model separately. 
Inspired by {\it Curriculum Learning} \cite{y_bengio_icml_2009_00}, 
we propose two training strategies in this work: Gradual Application of Enhanced Feature (GAEF) 
and Gradual Reduction of Enhancement Loss (GREL).
With GAEF, training initially starts using clean training features. Subsequently the portion of
outputs from the enhancement block gradually increases. With GREL, the Mean Squared Error
(MSE) loss for the enhanced output gradually reduces as training proceeds. In experimental results
on the LibriSpeech  {\tt test-clean}, {\tt test-other}
\cite{v_panayotov_icassp2015}, and the re-recorded far-field noisy test sets, the 
  NE-ASR model with GAEF-GREL training strategies
shows significantly better results than the conventional data augmentation
technique with relative improvements of 18.45 \%, 4.81 \%, and 12.36 \%
respectively. It has been very difficult to maintain the speech recognition
accuracy on clean speech with a MoCha-based model  when data augmentation is done using
the {\it acoustic simulator} to enhance far-field noisy performance \cite{c_kim_interspeech_2017_00}.
However, using the NE-ASR model with GAEF-GREL training strategies, the performance on
LibriSpeech {\tt test-clean} and {\tt test-other} sets is even substantially better
than the baseline MoCha model without noticeable training stability issues.
%
%
%
%
\bibliographystyle{IEEEtran}
\bibliography{common_bib_file}
\end{document}